# Non-equilibrium pathways to emergent polar supertextures

Vladimir A. Stoica[1,2,16,17], Tiannan Yang[1,16], Sujit Das[3,16], Yue Cao[4], Huaiyu Wang[1], Yuya Kubota[5,6], Cheng Dai[1], Hari Padmanabhan[1], Yusuke Sato[7], Anudeep Mangu[8], Quynh L. Nguyen[9,10], Zhan Zhang[2], Disha Talreja[1], Marc E. Zajac[2], Donald A. Walko[2], Anthony D. DiChiara[2], Shigeki Owada[5,6], Kohei Miyanishi[6], Kenji Tamasaku[5,6], Takahiro Sato[9], James M. Glownia[9], Vincent Esposito[9], Silke Nelson[9], Matthias C. Hoffmann[9], Richard D. Schaller[11], Aaron M. Lindenberg[8,10,12], Lane W. Martin[13,14], Ramamoorthy Ramesh[13,14,15], Iwao Matsuda[7], Diling Zhu[9], Long-Q. Chen[1], Haidan Wen[2,4,17], Venkatraman Gopalan[1,17], and John W. Freeland[2,17]

[1]Dept. of Materials Science and Engineering, Pennsylvania State University, University Park, PA 16802, USA
[2]Advanced Photon Source, Argonne National Laboratory, Lemont, IL 60439, USA
[3]Materials Research Centre, Indian Institute of Science, Bangalore, 560012, India
[4]Materials Science Division, Argonne National Laboratory, Lemont, IL 60439, USA
[5]Japan Synchrotron Radiation Research Institute, 1-1-1 Kouto, Sayo-cho, Sayo-gun, Hyogo 679-5198, Japan
[6]RIKEN SPring-8 Center, 1-1-1 Kouto, Sayo-cho, Sayo-gun, Hyogo 679-5148, Japan
[7]Institute for Solid State Physics, The University of Tokyo, Kashiwa, Chiba 277-8581, Japan
[8]Department of Materials Science and Engineering, Stanford University, Stanford, CA 94305, USA
[9]Linac Coherent Light Source, SLAC National Accelerator Laboratory, Menlo Park, CA 94025, USA
[10]Stanford PULSE Institute, SLAC National Accelerator Laboratory & Stanford University, Menlo Park, CA 94025, USA
[11]Center for Nanoscale Materials, Argonne National Laboratory, Lemont, IL 60439, USA
[12]Stanford Institute for Materials and Energy Sciences, SLAC National Accelerator Laboratory, Menlo Park, CA 94025, USA
[13]Department of Materials Science and Engineering, University of California, Berkeley, CA 94729, USA
[14]Materials Sciences Division, Lawrence Berkeley National Laboratory, Berkeley, CA 94720, USA
[15]Present address: Rice University, Houston, TX 77005.
[16]These authors contributed equally: V.A. Stoica, T. Yang, S. Das
[17]e-mail: vxs30@psu.edu, wen@anl.gov, vxg8@psu.edu, freeland@anl.gov

**Ultrafast stimuli can stabilize metastable states of matter inaccessible by equilibrium means. Establishing the spatiotemporal link between ultrafast excitation and metastability is crucial to understanding these phenomena. Here, we use single-shot optical-pump, X-ray-probe measurements to provide snapshots of the emergence of a persistent polar vortex supercrystal in a heterostructure that hosts a fine balance between built-in electrostatic and elastic frustrations by design. By perturbing this balance with photoinduced charges, a starting heterogenous mixture of polar phases disorders within a few picoseconds, resulting in a "soup" state composed of disordered ferroelectric and suppressed vortex orders. On the pico-to-nanosecond timescales, transient labyrinthine fluctuations form in this "soup" along with a recovering vortex order. On longer timescales, these fluctuations are progressively quenched by dynamical strain modulations, which drive the collective emergence of a single supercrystal phase. Our results, corroborated by dynamical phase-field modeling, reveal how ultrafast excitation of designer systems generates pathways for persistent metastability.**

There has been a tremendous interest in recent years in ultrafast optical control of solid-solid phase transformations by non-equilibrium means with an eye towards accessing transient or metastable phases and physics [1-3]. Excitation of condensed matter using short light pulses typically involves "melting" of charge [4], magnetic [5] and structural [6] orders. In contrast, ultrafast light can also be used to generate a new, transient (short-lived) order [7-13] and in a few cases, ultrafast light has created persistent (long-lived) metastable order away from the thermodynamic ground state [14-18]. Given the scarcity of examples and understanding of how these new types of ordered phases form through ultrafast excitation, it is natural to ask how one can achieve such metastable light-created states in a deterministic and thus abundant manner. A key to answering this question is to be able to directly observe, on relevant spatial and time scales, the precise pathways that a system takes from the moment of optical excitation to the creation and persistence of the metastable phase. If the metastable phase transformation is optically stimulated "one way" (i.e., persistent, and thus nonvolatile, until it is later erased by other means), then tracking large spatial and temporal scales becomes challenging and requires single-shot, pump-probe experiments that can directly record the order parameters involved in the transition [19-23]. Different mechanisms that can take place include coherent control of order conversion [1,2,9-11], stimulation of ultrafast disorder before a new order is formed [6], and ultrafast manipulation of intertwined orders near topological defects [24,25]. The relevant scales to investigate the role of charge, strain, heat, and other relevant order parameters can span many orders of magnitude in time and space.

As a model system, we focus on superlattices of PbTiO$_3$ (PTO) and SrTiO$_3$ (STO) that host a variety of extended polar textures such as supercrystals [16], vortices [26], skyrmions [27], and dipolar waves [28], collectively dubbed *polar supertextures*. These structures exist at the sub-angstrom-to-tens of nanometer length scales and are created by processes on femto-to-milli-seconds timescales. We start with a mixture of conventional ferroelectric (FE) (*i.e.*, $a_1/a_2$ twin structures) and more exotic polar vortices (V), exhibiting two length scales of periodically modulated structures: phase coexisting superdomains (~ 400 nm periodicity) containing FE and V within them (10-13 nm periodicity) [16,29]. Next, we alter the electrostatic boundary conditions of the superlattices with ultrafast optical excitation, and the system transforms into yet another novel, long-range-ordered phase called a vortex supercrystal (VSC) with a unit cell of $11 \times 30 \times 25$ nm$^3$, which is a persistent (*i.e.*, nonvolatile) state that after light stimulation has ceased [16]. Such three-dimensional phases were recently predicted to form under inhomogeneous charge-carrier redistribution [30], in contrast to other polar supercrystals observed under thermal equilibrium [31,32]. However, no experiments to date have revealed the mechanisms underlying the VSC supercrystal formation, their transition pathways, and the role of heterogeneous starting conditions. Furthermore, the coveted experimental access to VSC formation dynamics can inform prospects for persistent metastability more generally and expand existing ultrafast functionalities in heterogeneous quantum materials [13,15,18,24,25,33,34].

Here, we experimentally demonstrate that the nonvolatile phase-transition pathway to the metastable ordered VSC state occurs over multiple orders of magnitude in both time and length scales. This pathway develops by synergistic interplay between charge and structural heterogeneities that ultimately converge into the newly ordered phase by homogenization. Using femtosecond optical excitation and single-shot ultrafast X-ray diffraction, we uncover the complex evolution of the polar order from nano- to meso-spatial scales (10-400 nm) and over several orders in magnitude in time (0.1-10$^6$ ps) by tracking characteristic diffraction peaks to monitor a dynamical succession of phase transformations. The dynamics of satellite peaks connected to specific polar textures are indicated in **Fig. 1** and were obtained with an X-ray free electron laser (XFEL) experiment (**Extended Data Figs. 1-2 & Methods**). The results are summarized as follows.

Optical pulses with <100 fs duration, 3.1 eV energy (above the bandgap of PTO but below the bandgap of STO) and fluence of > 50 mJ/cm$^2$, generate charge carrier populations followed by their sub-ps migration to the PTO-STO interfaces. Within a few ps, the pristine FE- and V-phases take distinct paths to a fluctuating transient state formed under spatial charge carrier reorganization and thermalization of the system. The FE-phase is erased leaving behind disordered weakly nano-polar regions (dubbed D-FE). The V on the other hand converts to a metastable ferroelectric like $c^+/c^-$ stripe phase (up and down polarized nanoregions separated by walls) with suppressed vorticity formed by "unwinding" of the V under a reduced curl of polarization. The disordered-FE and the $c^+/c^-$ stripe phases, together are dubbed the "primordial soup," which subsequently forms a key intermediate link between the initial state and the supercrystal order. Over the 100 ps to a few nanoseconds timescale, a transient fluctuating labyrinthine phase, L, emerges, while the V order re-emerges and strengthens. At the same time, the original FE phase does not recover. The multi-phase heterogeneity evolves under a long-lived combination of charge carriers screening the depolarizing field and reduced mesoscopic ferroelastic strains at the phase boundaries. During film cooling via thermal diffusion into the substrate, the sequential phase transformations of intermediate phases culminate with their conversion to a single VSC ordered state, which starts to take over the intermediate phases at > 20 ns. These results are substantiated by dynamical phase-field modeling (DPFM) [35] that incorporates the spatial charge carrier redistribution [30] and extends the relaxational polarization kinetics to the strain and polarization dynamics for the modeling of polar vortex supercrystals [16,36,37].

To fully reveal the succession of phase transformations observed in experiments, we separately discuss the XFEL measurements over two temporal regimes, short (< 10 ps) timescale first and the long (> 10 ps) timescale next. The mixed-phase state responds promptly to the optical generation of a high density photocarrier population (~ 1-2 x10$^{21}$ cm$^3$) that is followed by a transient temperature jump of ~300 K (**Methods**) on a few ps timescale. **Figure 2a** depicts detector images near the 013 Bragg peak at select delay times, where the response of the V and FE phases are simultaneously monitored via characteristic satellite peaks. The intensity of the first-order satellite peak of V decreases by ~80 % within 0.5 ps, which is much faster than that measured during reversible dynamics [38], while the decrease of the FE satellite peak intensity is slower (~1 ps) (**Figs. 2b & Extended Data Fig. 3**). Since the width and peak position of the satellite peaks are not visibly changed at < 1 ps, this indicates that the peak intensity reduction is due to a change in the magnitude of the polarization rather than a strain change or alteration of the mesoscale structure [39]. This is further supported by the enhanced specular superlattice (SL) peak of V (marked in the inset, **Fig. 2a**), which is due to changes in the contribution of the polarization to the structure factor [40]. On the other hand, the strain determined from the position of the V and FE satellites develops on a longer timescale of 2-3 ps, which is close to a carrier-to-lattice thermalization time of ~ 4 ps obtained directly from calculations of electron-phonon scattering rates in PTO [41]. Additional support for the delayed strain response is obtained from the SL peak of the FE monitored by rocking around the Bragg condition (**Extended Data Fig. 4**), which confirms the unshifted peak at 1 ps and its strong shift of ~ 0.3% at 10 ps, stimulating a polarization reorientation toward the $z$-axis. Moreover, the opposite peak shift along $z$-axis for the V and FE phases suggests the simultaneous relaxation of strain difference along the $y$-axis [16,29], in the periodic direction of the phase mixture due to Poisson strain effects. To distinguish ultrafast polarization responses with electronic origins [42] from heat-driven depolarization [29,43] resulting from a sub-ps drop in the polarization, we examine in greater detail, the role of photocarrier excitation in the ultrafast response of the V- and FE-phases.

The dynamics of photocarriers are probed directly by transient-absorption (TA) spectroscopy in reversible dynamics regimes with lower pump fluence (< 10 mJ/cm$^2$). Using a white-light probe at energies below the bandgap absorption edges of both PTO and STO, TA probes transient charge-carrier populations evolving across the electronic bands of the SL following a dominant band-to-band optical pump absorption in PTO (Methods). The spectral response (**Fig. 2c**) indicates anomalous enhancement in the TA signals at ~1.17 eV and ~1.31 eV forming within the first 0.5 ps associated, respectively, with the electron and hole states that become separated at the PTO/STO interface; the known band offsets between STO and PTO are in good agreement with these measurements [44] (inset, **Fig. 2c**). The detailed temporal response at 1.17 eV shows a partial drop in the absorption at ~1 ps followed by a slow decay lasting into the nanosecond regime, but markedly different than that from the strain relaxation rate over tens of nanoseconds (**Extended Data Fig. 5**). The fast charge decay process can, for example, be assisted by bulk-photovoltaic effects [45] that provide high, nonequilibrium charge-carrier mobilities to enhance charge transfer at PTO/STO interfaces. DPFM simulations (**Figs. 2d & Extended Data Fig.6**) further confirm the migration and temporal response of the charge-carriers at interfaces.

The non-equilibrium changes in the electrostatic boundary conditions due to charge injection leads to a unique response that deviates from the thermodynamic phase diagram. While the FE-phase vanishes within ~10 ps, the V-phase is "unwound" and long-lived through the transient heating at ~ 600 K, albeit with a lower polarization. This is in stark contrast to the equilibrium response where the V-phase is suppressed at temperatures of ~ 500 K, while the FE state remains stable up to ~ 600 K [16,29,37], highlighting the unique conditions created by photocarrier separation at interfaces as the DPFM simulations demonstrate (**Fig. 2d**). The second-order V satellite is faint before the excitation (**Fig. 2a**), and it peaks at a different $q_z$ from the first-order satellite due to the vorticity of the polarization vector field. After optical excitation, however, the second-order peak at the same $q_z$ as the first-order peak becomes strong in 1.1 ps, suggesting a dramatic polar supertexture change driven by the electrostatic modification of the interfacial boundary conditions [16,30,37], which promotes a polarization reorientation along the vertical direction. In the V-phase, the polarization rotation generates local bending of atomic planes relative to the substrate interface, while if a polarization reorientation occurs, the bending deformation can be removed by "unwinding" of the vortex into the $c^+/c^-$ state, changing the intensity ratio of satellites. These sub-ps, non-equilibrium polarization responses are markedly faster than strain and thermal responses [42,43,46,47], as well as the expected thermalization timescale of ~ 4 ps [41], while a decoupled FE and V response is observed as well. Overall, the picosecond dynamics highlight how the combination of photocarriers and ultrafast heating generates the transient "primordial soup" composed of disordered FE and a metastable $c^+/c^-$ phase. This state with a mixture of ordered and disordered sub-systems is an essential starting point for the non-equilibrium pathway to the VSC state, as the system moves towards a new equilibrium.

Over longer timescales, we track the evolution of satellite peaks to monitor the order-disorder competition. In **Fig. 3a**, we present the detector images captured near the 002 Bragg peak that monitor the diffraction around the first- and second-order satellite peaks of the V-phase. On the scale of tens of picoseconds to nanoseconds, the $c^+/c^-$ state remains at a similar level of suppression until it starts to abruptly grow in intensity after a few ns, indicating the recovery of the V structure. Moreover, we see a new transient phase at > 100 ps as revealed by new broad satellites at a different $q_x$ and $q_z$ that were not present before the laser excitation (**Fig. 3a-b & Extended Data Fig. 7**); this is a new phase emerging in the disordered FE (D-FE) region. Using DPFM, we identify this new phase to exhibit labyrinthine fluctuations [48], dubbed the L-phase.

In the ns regime, a transient picture for metastable phase coexistence can be accordingly established, where the spatial regions encompassing the V-phase remain stable, albeit with a reduced transient polarization, while the FE regions are replaced by the L-phase.

At ~5 ns, the L and V peaks start to grow strongly in intensity at the same rate, while a delayed onset of the VSC peak growth is observed at a characteristic time of $\tau \approx 20$ ns (**Fig. 3c**). The analysis of dynamic SL peak position of the V-phase and satellite peak widths for L and V along the in-plane direction (**Fig. 3d**) shows two distinct dynamical regimes indicated by yellow and blue background colors. At times before $\tau$, despite the simultaneous peak growth for L- and V-phases, the peak width of the former strongly narrows, while the peak width of the latter remains unchanged, respectively. This indicates decoupled dynamics of coexisting transient phases before $\tau$. Near $\tau$, after the peak width of L-phase approaches the width of the V-phase, the smaller periodicity of the L-phase starts to suddenly increase and approach the one of the V-phase. This relative difference between periodicities of different phases suggests the formation of dislocation dipole defects [49] at the boundary of the two stripe lattices, which is consistent with their annihilation at times $> \tau$ when the two-phase system merges into the VSC phase. The onset of VSC emergence drives the necessary periodicity adjustment in the L-phase to match the VSC periodicities, which allows for spatial homogenization. This picture is supported by the concurrent second-order satellite peak development in the L-phase (**Fig. 3b & Extended Data Fig. 7**), indicative of a transition to long-range ordering at the onset ($\tau$) of VSC formation. Furthermore, a strong non-monotonic peak shift, resembling an overshoot, develops along $q_z$ in the V-phase at $\tau$ as well, which is assigned to the strain-mediated coupling of nanoscale (polar supertexture length scale) and mesoscale (superdomain length scale) orders as the VSC is formed. We observe this via the out-of-plane strain change that starts to decrease again at $> 300$ ns, following the increase started at 20 ns, to ultimately converge toward the characteristic strain values of the VSC at 1 μs. Overall, these experimental observations conclude that a dynamical regime of strain criticality occurs at $\sim \tau$, where elastic interactions at the superdomain boundaries synchronize the simultaneous evolution of the L- and V-phases into a single VSC-phase. These findings highlight the critical role played by mesoscale strain interactions in the dynamical realm of stabilizing novel metastable orders by non-equilibrium routes.

The direct experimental observation of the mesoscale-phase evolution is accurately described by DPFM (**Methods**), allowing us to monitor the coupling of charge carrier dynamics and thermal relaxation rate (**Extended Data Fig. 5**) with spatially resolved polarization and strain [30,35]. To build a picture for the real space evolution of the phases, we map the spatial dependence of the simulated volumes (**Fig. 4a**), showing the distributions of polarization, electron concentration, polarization vorticity, and strain in the ground state and at representative time delays after optical excitation. Using these maps, the real-space evolution of the different phases becomes evident. The initial V region indeed unwinds into a $c^+/c^-$ structure before reforming on the nanosecond time scale as suggested by the experiment. At the same time, the initially disordered FE region is replaced with the L-phase at ~ 5 ns and, in turn, both regions evolve into the VSC-phase at 50 ns driven by ferroelastic strain interplay at the phase boundary. Most of the available space is filled with VSC, while a small remnant-phase becomes trapped within and requires longer times for phase conversion. In further support of the mesoscopic interplay between phases, a simulation starting from a pure-FE phase (**Extended Data Fig. 8**) captured the transient L-phase instability as well, while the VSC transition remained absent in this case. Accordingly, we can conclude that DPFM captures the essential nature of the phase transition observed in the experiment.

For a full overview of the dynamics (**Fig. 4b**), we present representative parameters extracted from simulations on extended timescales. The spatially averaged dynamics of the polarization components is compared with spatially resolved dynamics of polarization vorticity ($\nabla \times P_y$), shear ($\varepsilon_{23}$), and normal ($\varepsilon_{33}$) strains. Because the initial V-phase does not possess $P_y$, at short-time delays, $P_y$ directly correlates with the initial FE-phase, while at longer-time delays $P_y$ tracks the emergence of the VSC-phase, correlating with the out-of-plane to in-plane reorientation of polarization. Simultaneously, the magnitude of vertical polarization, $P_z$, a component absent in the FE-phase, tracks the order parameter of the V- and L-phases and their conversion to VSC. The $\nabla \times P_y$ and $\varepsilon_{23}$ both track the evolution of polarization winding with time, decreasing first within the first 10ps as the V phase unwinds due to charge injection, and increasing on nanosecond time scales as V reemerges and the VSC forms. The evolution of both the $\varepsilon_{23}$ and the $\varepsilon_{33}$ indicate a rapid strain homogenization of the initial phases (indicated by regions 1 and 2) within 10 ps and the "strain locking" between those regions subsequently all the way up to VSC formation. The simulated relative difference in $\varepsilon_{33}$ strain within phases tracks well with the corresponding $\varepsilon_{33}$ differences between the FE-, V-, and VSC-phases observed in the experiments. Furthermore, the non-monotonous overshoot of ~ 0.1% in $\varepsilon_{33}$ (**Fig. 4b**) between 3-30 ns, generated by polarization reorientation to the *y*-direction followed by walls forming between the in-plane and out-of-plane polarized regions of the VSC, is consistent with the experimental observation (**Fig. 3d**). Overall, the model for the transformation dynamics tracks very well with the experiment, which demonstrates its capability in capturing the essential aspects of this nonvolatile phase transition created from charge-carriers and thermal interplay directly impacting the strain and polar supertexture evolution. On short time scales, the FE-phase is largely erased by the temperature rise while the V-phase is stabilized by the charge separation at interfaces (*i.e.*, holes and electrons residing on the PTO and STO sides of the interface, respectively) that screens the depolarizing field and enables polarization reorientations. The charge separation at interfaces is long-lived and cooperative with the lattice cooling resulting in anisotropic strain changes that push the system into a persistent VSC state.

The dynamical trajectory of successive phase transformations in a three-dimensional parameter space is summarized (**Fig. 5**), which compares the local strain and charge reorganization of the starting phases along the pathway of evolution. The shear strain is generated at the intralayer boundary between nanodomains with different polar textures and is due to geometrical mismatch of lattice deformations linked to abrupt polarization reorientations across the respective boundaries. A larger shear strain also reflects a tendency for the FE polarization to continuously "curl", that is, be oriented neither completely in the film plane, nor completely perpendicular to it, but in an intermediate direction as in the V phase. The resulting structure of the VSC vertically doubles the heteroepitaxial periodicity [16,37] and is assisted by cooperative ordered tilts due to shear strain modulations inside the superlattice [36]. The electron concentration reflects the optically excited carriers and their subsequent relaxation. After optical excitation, the shear strain and charge difference between the V and the FE regions is quickly suppressed in a few ps (dotted lines) during the corresponding timescale of electronic non-equilibrium probed in experiments.

Our results on polar supertextures elucidate a non-equilibrium pathway by which an optical pulse can create a new ordered phase starting from a heterogenous ground state in superlattice structures, where photo-induced charges perturb the fine balance between electrostatic and elastic energies that is built-in during growth. The irreversible phase transition that ensues is probed using single shot ultrafast X-ray probes over seven orders of magnitude in time, which reveals an evolving reconfiguration of polar topology in synergy with the ferroelastic strain progression at the phase boundary. This study thus provides a template for tracking such novel phases and their competition via state-of-the-art single-shot diffraction experiments. In combination with dynamical mesoscale theory, we establish a paradigm to generate other undiscovered metastable polar super textures by control of light characteristics harnessed in heterostructures. While phase heterogeneity and competition are key universal properties of emergent systems, a broad opportunity exists to create designer heterogeneous phases, and to control their mutual competition and cooperativity with external stimuli. Enhancing order with light remains a rare phenomenon and when it occurs on ultrafast timescales, it is even more unexpected. Moving beyond polar order, this approach can be expanded to explore emergent electronic and magnetic phases, where phase competition is an essential ingredient, and the modalities for how light couples with matter can be even richer [1,3,50]. Driving the phase heterogeneity, ubiquitous in quantum materials [13,15,24,25,33,34], into hitherto unexplored states offers tremendous opportunities to harness novel functionalities via non-equilibrium pathways.

## Methods

**Single-shot optical-pump, ultrafast X-ray diffraction measurements**

The measurements were performed at the BL3 beamline of SACLA [51] X-ray free electron laser (XFEL) and X-ray Pump Probe beamline of the Linac Coherent Light Source (LCLS) at room temperature. Monochromatized hard X-ray pulses at 9.5 keV at LCLS and 10 keV at SACLA were focused to < 40 μm beam size with beryllium compound refractive lenses, while the optical pump beam at 400 nm had a beam size of ~ 100 μm. A diamond (1 1 1) monocromator with an estimated bandwidth of 0.7 eV around the central energy was used at LCLS. The X-ray beam was attenuated to avoid sample damage and detector saturation. At SACLA XFEL, the X-ray pulses had a duration of <7 fs and a repetition rate of 30 Hz in continuous mode of operation. The LCLS delivered X-ray pulses with a pulse duration of ~ 40 fs at a repetition rate of 30 Hz (reduced by pulse picker), while the continuous mode of operation was 120 Hz. At LCLS, single X-ray pulses were pulse picked and delivered on demand, synchronous with single laser pulses picked up by Pockels cell trigger. The sample was mounted on a multi-circle goniometer in the horizontal scattering geometry. Scattered X-rays were captured shot-by-shot using a high dynamic-range area detector (multi-port CCD [52] at SACLA and Jungfrau 1M at LCLS), positioned at a radial detector distance of 400-600 mm from the sample.

A Ti:sapphire femtosecond laser system synchronized to the XFEL was used to generate 400-nm ultrafast laser pulses (at the second harmonic of the fundamental laser wavelength) with variable laser fluence of 50-150 mJ/cm$^2$. A measurement protocol recorded the diffraction patterns from 20-30 X-ray pulses (acquired at 30 Hz rate) before the sample irradiation with a single optical pulse, followed by a pair of successive optical and X-ray pulses for time dependent determination of the dynamics. At the end of the measurements in each spatial location of the sample, another 20-30 X-ray pulses (acquired at 30 Hz rate) recorded the diffraction patterns after the sample irradiation with the single optical pulse, confirming the nonvolatile switching of the phases from a V+FE mixture to the VSC phase.

During single-shot measurements, the optical pulse duration was < 100 fs, while the incident laser fluence was above the threshold for single-shot switching of the phases [16]. Before the experiment on each sample, the sample tilt was pre-aligned using a tilt stage to remain parallel within < 10 μm with a motorized stage translating the sample parallel to its surface. A 10× microscope objective with high sensitivity to focus condition was mounted on a charge coupled device (CCD) for optical imaging during the sample translation.

During the experiment, after each laser shot on the sample and collection of diffraction data, the sample was translated to different fresh (pristine) spatial locations. Since the laser induced phase transition could be reversed by heating [16], some of the samples were reused in experiment after reversing to pristine phases after heating to > 650 K and cooling back to room temperature. The corresponding X-ray scattering patterns collected around selected diffraction conditions were acquired as a function of delay between the optical pump and X-ray probe pulses with a temporal resolution of < 1 ps and with variable time delay on fs-ms timescales. To normalize statistical variations of time-resolved data (see methods) from each spot on the sample, complimentary reference diffraction patterns were recorded before and after the dynamical transient spaced out in 30 ms increments given by X-ray pulse separation from XFEL. The raw detector images were accordingly grouped into laser before/on/after data sets and normalized against the single X-ray pulse intensity monitor, $I_0$. The fitting of the satellite diffraction peaks used Voigt function profiles and linear background to extract peak amplitudes, positions, and widths as a function of time delay.

**Synchrotron characterization of the samples**

The dominant absorption of the 3.1 eV pump excitation used in our work is in $PbTiO_3$ layers of the superlattice as was shown in prior measurements in $PbTiO_3$ films [42], while the larger bandgap of $SrTiO_3$ at 3.2 eV [53] suggests a weaker absorption in this layer. This is confirmed in control thin film samples with ~ 50 nm thickness; we compared the thin-film peak shifts, which measures the transient strain at beamline 7-IDC of APS with an excitation fluence of < 10 mJ/cm$^2$. Only the $PbTiO_3$ reference film gave a measurable transient strain response, while $SrTiO_3$ reference film did not provide a measurable response, confirming the primary optical absorption in the $PbTiO_3$ layers. Regarding the strength of photocarrier generation, we use the optical absorption coefficient from prior reports [16,42] to estimate an initial carrier concentration ~ 1-2x10$^{21}$ cm$^{-3}$ in the single-shot measurements, while the charge carrier relaxation rate is obtained from transient absorption measurements (Extended Data Fig. 5a-b). Likewise, we obtain an estimate for temperature change from the previously determined thermal expansion coefficient [29,38] combined with FE phase peaks shifts that follow a linear fluence dependency at 600 ps and at < 12 mJ/cm$^3$, below the onset of nonvolatile photoinduced changes to the sample (Extended Data Fig. 5c). We extrapolate linearly these results to the high fluence used for single-shot probing of nonvolatile dynamics at XFEL. The FE strain measurement along the vertical direction of the sample is probed along the non-polar axis of the structure and is expected to follow the thermal expansion coefficients in a directly proportional fashion compared to the strain components along to polarization direction. Using: a) the thermoelastic model of thermo-elastic lattice expansion of a film attached to unexcited substrate [54], where the in-plane thermal expansion is impeded at short time delays of < 1 ns, b) a linear extrapolation from Extended Data Fig. 5c to high fluence regime, and c) the reported Poisson ratios of the PTO [55] and STO [56], the initial temperature jump at ~ 100 mJ/cm$^2$ is estimated to be ~ 300 K in the PTO-STO superlattice samples. At the APS, the laser pulse with a fluence of ~ 100 mJ/cm$^2$ delivered at normal incidence exceeded levels needed to fully switch the pristine phases to VSC, while the optical pulse duration, the pump wavelength, and incident angle of the sample (changing the absorbed fluence) can modify the fluence levels to achieve the full conversion to VSC [16]. To match the APS observations at XFEL and account for variations of absorbed fluence as a function of incident beam angle combined with pulse duration, the fluence values at XFEL are adjusted based on the nonvolatile response detected in the "after" reference diffraction patterns (**Fig. S1d**), which confirm the suppression of the FE-phase and the fully developed VSC peak. Moreover, to compare results obtained for 002 and 013 peaks in single-shot XFEL measurements, we also adjusted the fluence to match the strength of V satellite intensity reduction at the ps timescale combined with the permanent disappearance of the FE peak in these two peaks of interest.

**Sample growth and characterization**

The (PbTiO$_3$)$_{16}$/(SrTiO$_3$)$_{16}$ superlattice samples on (110) DyScO$_3$ substrates were fabricated using pulsed laser assisted deposition. Structural sample characterization and preliminary time-resolved measurements employed X-ray synchrotron diffraction at the Advanced Photon Source, Argonne National Laboratory, using beamlines 33-ID-C and 7-ID-C, respectively. Three-dimensional reciprocal space mapping (RSM) was used to confirm the as-grown (pristine) quality of the samples, confirming a mix of polar vortex and $a_1/a_2$ twin FE structures in the superlattices. The FE structures exhibit conventional $a_1/a_2$ twin stripe domains with 1.6% smaller out-of-plane lattice constant compared with the vortex structures at room temperature, which allows us to clearly distinguish the peaks of interest.

**Phase-field simulations**

We simulate the structural, carrier, and temperature evolution during optical excitation of the sample using a dynamical phase-field model. The system is described by the spatial distribution of a set of independent variables, including the polarization **P**, the mechanical displacement **u**, the free electron and hole concentrations $n$ and $p$, and the electric potential $\Phi$. The time-dependent evolution of these variables is simulated by solving a set of governing equations of the system, including the polarization dynamics equation, the elastodynamics equation, the electron and hole transport equations, and the electrostatic Poisson equation, written as [30,35,57]

$$\boldsymbol{\mu}\ddot{\mathbf{P}} + \boldsymbol{\gamma}\dot{\mathbf{P}} = -\frac{\delta F}{\delta \mathbf{P}}, \tag{1}$$

$$\rho\ddot{\mathbf{u}} = \nabla \cdot (\boldsymbol{\sigma} + \beta\dot{\boldsymbol{\sigma}}), \tag{2}$$

$$\dot{n} = -\nabla \cdot \mathbf{j}_e + S_e + R, \tag{3}$$

$$\dot{p} = -\nabla \cdot \mathbf{j}_h + S_h + R, \tag{4}$$

$$\nabla \cdot \left(-\kappa_0 \boldsymbol{\kappa}^b \nabla \Phi + \mathbf{P}\right) = e(p - n). \tag{5}$$

In equation (1), $\boldsymbol{\mu}$ and $\boldsymbol{\gamma}$ are the mass and damping coefficients of the polarization and $F$ is the total free energy of the system with expression provided in the next section. In equation (2), $\rho$ and $\beta$ are the material mass density and the stiffness damping coefficient, respectively, and $\boldsymbol{\sigma}$ is the stress field. In equations (3) and (4), $\mathbf{j}_e$ and $\mathbf{j}_h$ are the net fluxes of free electrons and holes, respectively, $S_e$ and $S_h$ are carrier injection rates from external sources, and $R$ is the generation rate of electron-hole pairs, which includes contributions from intrinsic electron-hole creation/recombination and light-induced excitation. In equation (5), $\kappa_0$ and $\boldsymbol{\kappa}^b$ are the vacuum permittivity and background dielectric constant, respectively, and $e$ is the elementary charge.

The simulation system of the PbTiO$_3$/SrTiO$_3$ superlattice contains 4 PbTiO$_3$ layers and 4 SrTiO$_3$ layers, where each layer has a thickness of 4.8 nm, equivalent to 12-unit cell lengths of the perovskite crystal lattice. The in-plane dimensions of the system are chosen as 160 nm × 160 nm. Three-dimensional periodic boundary conditions are employed for polarization, strain, carrier concentrations, and electric potential. The spatial average of the out-of-plane stress components $\sigma_{13}$, $\sigma_{23}$, and $\sigma_{33}$ is fixed at zero for describing an out-of-plane stress-free condition, while the spatial average of the in-plane strain components $\varepsilon_{11}$, $\varepsilon_{22}$, and $\varepsilon_{12}$ are calculated from the difference between the average bulk lattice parameters of the superlattice relative to the substrate lattice parameters.

The pristine structure of a mixture of $a_1/a_2$ and vortex domains is first obtained by evolving the system to the equilibrium state under room temperature without considering the charge carriers. In this step, an optimal anisotropy of substrate lattice is used to match experimental observations of FE + V phase mixture in the ground state [16,29]. Next, the superimposed carrier dynamics in the ground state are modeled by fitting the relaxation times of the transient absorption measurements to make assumptions in the model, while a dynamic temperature profile can be created from a superimposed value of electron-phonon coupling time and strain relaxation rate observed experimentally. The optical excitation is modeled by charge injection that is initially uniformly distributed but allowed to spatially evolve. The polarization and strain distribution of the pristine state under charge-carrier concentration of 0 as the initial condition (at t=-1 ps) is followed the above-bandgap optical excitation of the nanostructure by considering:

A) the light-induced instant generation of free carriers, with a time-dependent generation rate $R_{light}$ following a Gaussian function describing the application of a light pulse peaking at $t_0 = 0$, written as

$$R_{light} = R_0 \exp\left[-(t-t_0)^2/\tau_p^2\right], \qquad (6)$$

with $\tau_p = 0.1$ ps. $R_0$ is given by the total concentration of electrons generated across the light pulse duration, which is estimated as $10^{21}$ cm$^{-3}$ [16,42]; and

B) a fast rise of the temperature to $T_m = 600$ K with a time constant of $\tau_r = 1$ ps describing the thermal effect of the light pulse through electron-phonon coupling after the light application at $t_0 = 0$, followed by an exponential decay to the room temperature with a time constant of $\tau_d = 20$ ns, written as

$$T = T_0 + (T_m - T_0)\{1 - \exp[-(t-t_0)/\tau_r]\}\exp[-(t-t_0)/\tau_d]. \qquad (7)$$

By numerically solving equations (1-5), we obtain the spatiotemporal evolution of the coupled electron-lattice processes during optical excitation of the PbTiO$_3$/SrTiO$_3$ superlattice.

The material constants for PbTiO$_3$ used in the phase-field simulations include the following [30]. The polarization mass and damping coefficients are $\mu_{11} = 7.5 \times 10^{-17}$ J m s$^2$ C$^{-2}$ and $\gamma_{11} = 5.9 \times 10^{-6}$ J m s C$^{-2}$, respectively [37]. The material mass density and the stiffness damping coefficients are $\rho = 7.5 \times 10^3$ kg m$^{-3}$ and $\beta = 6 \times 10^{-12}$ s, respectively. The background dielectric constant is taken as $\kappa_{11}^b = 20$. The Landau coefficients are $a_1 = (T/K - 752) \times 3.8 \times 10^5$ J m C$^{-2}$, $a_{11} = -0.725 \times 10^8$ J m$^5$C$^{-4}$, $a_{12} = 7.50 \times 10^8$ J m$^5$C$^{-4}$, $a_{111} = 2.606 \times 10^8$ J m$^9$C$^{-6}$, $a_{112} = 6.10 \times 10^8$ J m$^9$C$^{-6}$, and $a_{123} = -37.0 \times 10^8$ J m$^9$C$^{-6}$ [58]. The gradient energy coefficients are $g_{11} = 1.04 \times 10^{-10}$ J m$^3$C$^{-2}$, $g_{12} = 0$, and $g_{44} = 0.52 \times 10^{-10}$ J m$^3$C$^{-2}$. The elastic stiffness is $c_{11} = 80 \times 10^{11}$ J m$^{-3}$, $c_{12} = 0.80 \times 10^{11}$ J m$^{-3}$, and $c_{44} = 1.10 \times 10^{11}$ J m$^{-3}$ [58]. The electrostrictive coefficients are $Q_{11} = 0.089$ m$^4$C$^{-2}$, $Q_{12} = -0.026$ m$^4$C$^{-2}$, and $Q_{44} = 0.03375$ m$^4$C$^{-2}$ [59]. The band gap is $E_g = 3.0$ eV, and the effective density of states coefficients for valence and conduction bands under a parabolic approximation [35] are $K_{DV} = K_{DC} = 4 \times 10^{28}$ m$^{-3}$eV$^{-1.5}$ [60]. The electron and hole mobility is $M_e = M_h = 5 \times 10^{-4}$ m$^2$s$^{-1}$V$^{-1}$ and the electron-hole relaxation coefficient is $K_{eh} = 5 \times 10^{-16}$ m$^3$s$^{-1}$, which are within the range of existing reports [61,62] and tuned to reproduce the experimental carrier relaxation timescales in the present work.

For simplicity of the simulations, most of the material constants for SrTiO$_3$ are taken the same as those of PbTiO$_3$, except for the Landau coefficients, the electrostrictive coefficients, and the band gap. The Landau coefficients and the electrostrictive coefficients are taken as $a_1 = (\coth(54K/T) - 1.06) \times 4.05 \times 10^7$ J m C$^{-2}$, $a_{11} = 1.70 \times 10^9$ J m$^5$C$^{-4}$, $a_{12} = 3.92 \times 10^9$ J m$^5$C$^{-4}$, and $a_{111} = a_{112} = a_{123} = 0$, $Q_{11} = 0.0457$ m$^4$C$^{-2}$, $Q_{12} = -0.0135$ m$^4$C$^{-2}$, and $Q_{44} = 0.0096$ m$^4$C$^{-2}$ [63]. The band gap is taken as $E_g = 3.2$ eV for SrTiO$_3$ and the top of the valence band in SrTiO$_3$ is ~ 1.2 eV lower than that of PbTiO$_3$ [64].

## Data availability

All data supporting the findings of this study are available within the paper and/or are available from the authors upon reasonable request.

## Acknowledgements

This work is primarily supported by U.S. Department of Energy, Office of Science, Office of Basic Energy Sciences, under Award Number DE-SC-0012375 for diffraction and simulation-based study of the materials. R.R. and S.D. acknowledge support from the Office of Basic Energy Sciences, US Department of Energy (DE-AC02-05CH11231). L.W.M. and R.R. also acknowledge partial support of the Army Research Office under the ETHOS MURI via cooperative agreement W911NF-21-2-0162 for the development of superlattice structures. T.Y. and L.Q.C. also acknowledge partial support as part of the Computational Materials Sciences Program funded by the U.S. Department of Energy, Office of Science, Basic Energy Sciences, under Award No. DE-SC0020145. Q.L.N. acknowledges support from the Bloch Fellowship in Quantum Science and Engineering by the Stanford-SLAC Quantum Fundamentals, Architectures and Machines Initiative. Y.C. and H.Wen acknowledge the support by U.S. Department of Energy, Office of Science, Office of Basic Energy Sciences, Materials Sciences and Engineering Division. Use of the Linac Coherent Light Source (LCLS), SLAC National Accelerator Laboratory, is supported by the U.S. Department of Energy, Office of Science, Office of Basic Energy Sciences under Contract No. DE-AC02-76SF00515. The XFEL experiments were also performed at the BL3-EH2 of SACLA with the approval of the Japan Synchrotron Radiation Research Institute (JASRI) (Proposal No. 2019B8019). This research used in part resources of the Advanced Photon Source, a U.S. Department of Energy (DOE) Office of Science User Facility operated for the DOE Office of Science by Argonne National Laboratory under Contract No. DE-AC02-06CH11357, with data collected at X-ray Science Division beamlines 7ID-C and 33ID-B.

**Contributions**

V.A.S., J.W.F, H.Wen and Y.C. conceived the experimental design in consultation with Y.K., I.M., A.M.L. and D.Z. V.A.S., J.W.F., V.G. and H.Wen developed the central concepts of the research. V.A.S., A.D.D., H.Wen, D.A.W., H.P., and D.T. demonstrated the experimental proof of concept at synchrotron. S.D. synthesized the samples and performed basic laboratory X-ray diffraction with support from R.R. and L.W.M. V.A.S., Z.Z., H.Wen, H.P., D.T. and M.E.Z. conducted samples screening using synchrotron X-ray diffraction. V.A.S. carried out the reciprocal space map analysis of these samples with support from J.W.F, H.Wen, and Z.Z. T.Y. and L.Q.C. developed the phase-field modeling platform and carried out the simulations with assistance from C.D. T.Y. analyzed phase-field modelling results under L.Q.C. and C.D. guidance, and in close collaboration with V.A.S., V.G. and J.W.F. Y.K., D.Z., S.O., K.M., K.T., T.S., Q.L.N., J.W.G., V.E., S.N. and M.C.H. provided experiment control solutions at X-ray free-electron laser facilities. V.A.S., Y.C., H.Wang, Y.K., H.P., Y.S., A.M., Q.L.N., S.O., K.M., K.T., T.S., J.W.G., V.E., S.N., M.C.H., A.M.L., I.M., D.Z., H.Wen, V.G. and J.W.F. conduced the X-ray free-electron laser experiments. J.W.F., Y.C., H.Wen and V.A.S. developed the python-based scripts for the experimental data analysis. R.D.S. conducted the transient absorption (optical) spectroscopy measurements. V.A.S. conducted the experimental data analysis in close collaboration with J.W.F., V.G., T.Y., H.Wen, L.Q.C., Y.C., A.M.L., L.W.M., H.Wang, C.D. and R.D.S. V.A.S., J.W.F, V.G., L.W.M. and T.Y. wrote the manuscript with suggestions from the other authors. All authors contributed to the discussion and the final version of the manuscript.

# Figures

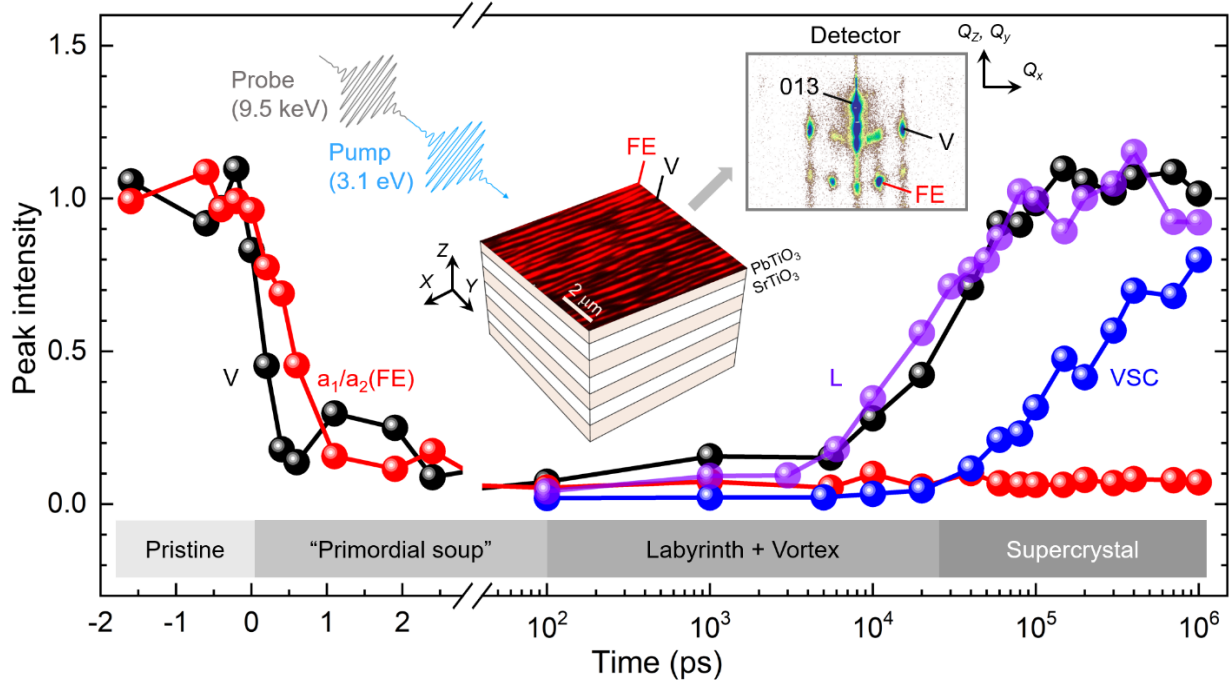

**Fig. 1 Single-shot probe of nonvolatile dynamics of vortex supercrystal formation.** Time resolved X-ray scattering measurements of nonvolatile phase switching dynamics in a $[(SrTiO_3)_{16}/(PbTiO_3)_{16}]_{7.5}$ superlattice sample with X-ray (probe) and laser (pump) beams in collinear geometry. The scattering plane is the y-z plane. Twin $a_1/a_2$ ferroelectric (FE) and polar vortex (V) phase coexistence in the pristine state is imaged using dark field X-ray imaging (XRIM) in Bragg diffraction condition shown in the inset, which captures the steady-state quasi-periodic mixing of phases generating superdomains along *y*-direction with ~ 400 nm periodicity. In dynamical measurements, the diffraction conditions are selected for the observation of pristine phases (i.e., FE and V) and vortex supercrystal (VSC) (see Extended Data Figs. 1-2 for additional details). The detector image captures the temporal evolution of peaks corresponding to in-plane periodic structures of 12.5 nm and 11 nm, for FE and V phases, respectively. The resulting peak intensity dynamics are plotted for comparison, where a succession of sequential transformations is indicated. After the initial collapse of the V and FE orders on ~ 1 ps timescale and thermalization in 2-4 ps, a "primordial soup" of disordered ferroelectric (D-FE) and metastable vortex (V) forms in the intermediate stages, followed by nucleation and growth of a labyrinthine (L) phase that correlates with the recovery of the V phase on a few nanoseconds time. The intermediate L and V phases undergo simultaneous conversion to the VSC supercrystal phase via nucleation and growth at > 20 ns.

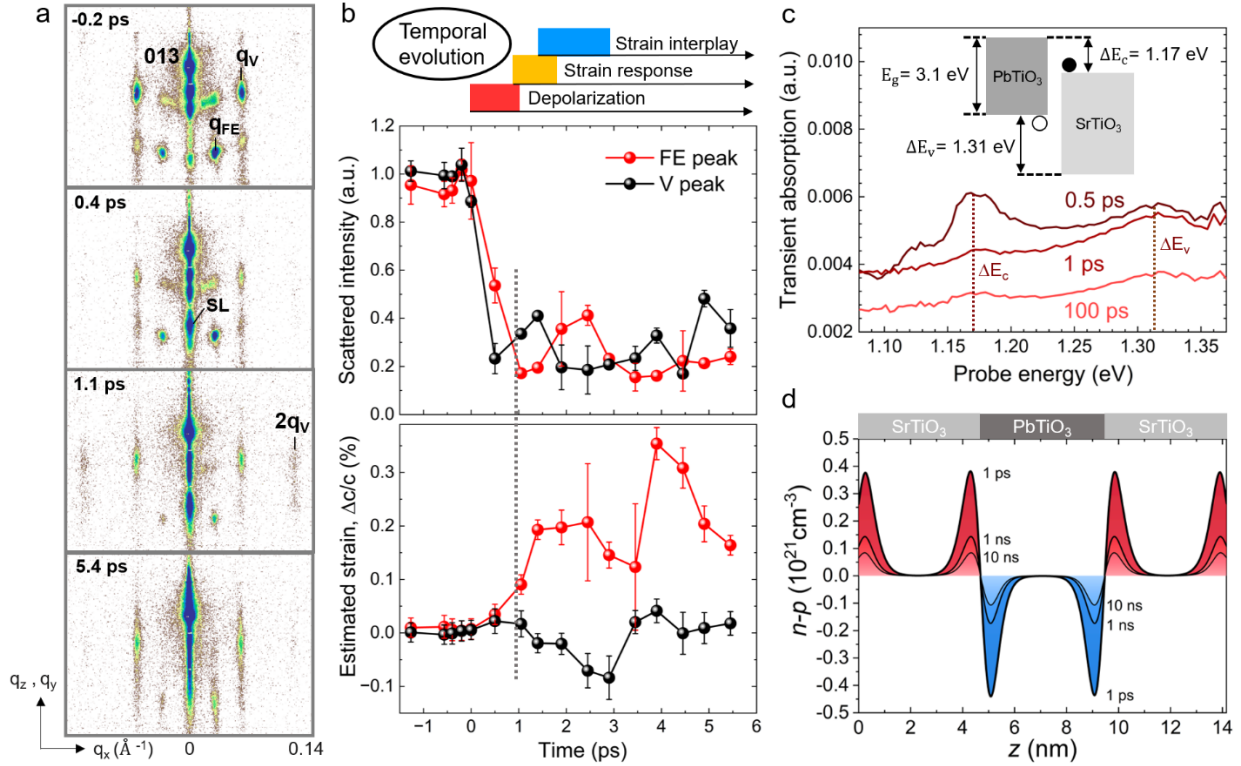

**Fig. 2 Non-equilibrium polarization dynamics correlated with interfacial charge separation in a [(SrTiO$_3$)$_{16}$/PbTiO$_3$)$_{16}$]$_{7.5}$ superlattice. a.** Detector images are shown in false color logarithmic scale and captured near the 013 Bragg diffraction condition (Extended Data Fig. 1b) at selected time delays after laser excitation (additional time delays shown in Extended Data Fig. 3). The main satellite peaks of V- and FE-phases, corresponding to their respective in-plane periodic order, are indicated. V and FE satellite peaks are suppressed at different time delays; whereas V is strongly diminished by 0.4 ps, the FE is diminished by 1.1 ps and broadens later (5.4 ps) along q$_z$. The emergence of second order V peak follows suit at 1.1 ps while an initially extinct superlattice (SL) peak corresponding to the V phase emerges at 0.4 ps and persists on the ps timescale. **b.** The temporal evolution of V and FE peaks indicates a succession of processes derived from the intensity and peak shift dynamics measured at the detector. The vertical line shows that peak intensity collapse in both phases is faster than strain dynamics measured by peak shifts. **c.** Reversible transient absorption (TA) spectroscopy measurements of the sample at small pump fluence of < 10 mJ/cm$^2$, indicating dynamic peak enhancements at 1.17 eV and 1.31 eV that are attributed to band offsets at the superlattice interfaces [41] and electron-hole separation across PTO-STO SL interfaces as shown in the inset. **d.** The temporal evolution of electron and hole distribution, calculated with DPFM, reveals electron abondance in STO and hole abundance in PTO near their interfaces, which agrees with the electronic band diagram in c.

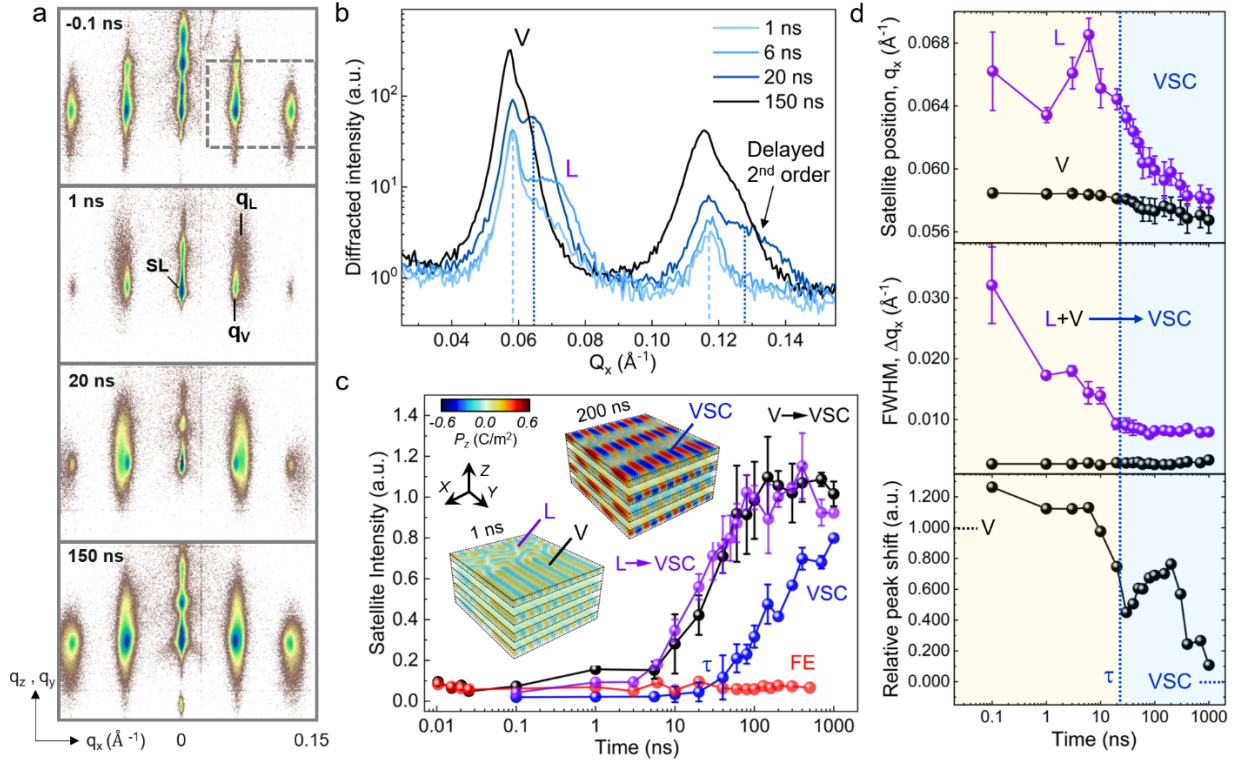

**Fig. 3 Distinguishing spatially dependent dynamics of transient phases driving the formation of VSC in [(SrTiO$_3$)$_{16}$/(PbTiO$_3$)$_{16}$]$_{7.5}$ superlattices. a.** Detector images acquired near 002 Bragg diffraction condition (Extended Data Fig. 1a) capturing a few time delays after laser excitation (detailed temporal dependence is shown in Extended Data Fig. 7). The scattering plane is the y-z plane. A transient broad peak assigned to labyrinthine (L) phase (marked at q$_L$) takes over the original FE-phase in the respective regions of the pristine sample. **b.** The temporal evolution of peak profiles, integrated inside the dashed grey box shown in **a**, shows first and second order satellite peaks of the V- and L- phases, which are marked by the vertical dashed and dotted lines, respectively. A second order replica of the L satellite peak appears abruptly at > 20 ns, indicating the L to stripe-like order transition with long range order. **c.** A summary of a few representative peaks normalized to reference detector images, showing a strong correlation between V and L peaks from ns to µs timescales. The FE peak stays suppressed and does not recover. The direct space representations of the transient fluctuating L- and V- phases at 1 ns and the full development of VSC-phase at 200 ns from the DPFM simulations are shown for comparison in the inset. **d.** The temporal evolution of peak positions and their full width at the half maximum (FWHM) is extracted from peak profile fitting along the x direction and is plotted in the top two panels for V- and L-phases. The bottom panel shows the vertical peak shift of SL peak marked in **a** and measuring the q$_z$ changes of the V-phase, which are proportional to the normal strain, $\varepsilon_{33}$. The peak shift is rescaled relative to peak position before laser excitation (labeled 1 on the *y*-axis), which is the strain state of the pristine V-phase, while the peak position in the final state (labeled 0 on the *y*-axis) corresponds to the strain state of VSC. A non-monotonic strain change versus time, including a strain "overshoot" (see text) starting at ~ 30 ns, captures the dynamics of phase conversion to VSC. The colored regions, separated by the characteristic time, τ, highlight two distinct dynamical regimes: transient phase L fluctuations are detected in the yellow region, while the formation of the VSC-phase occurs in the blue region (see text for the description).

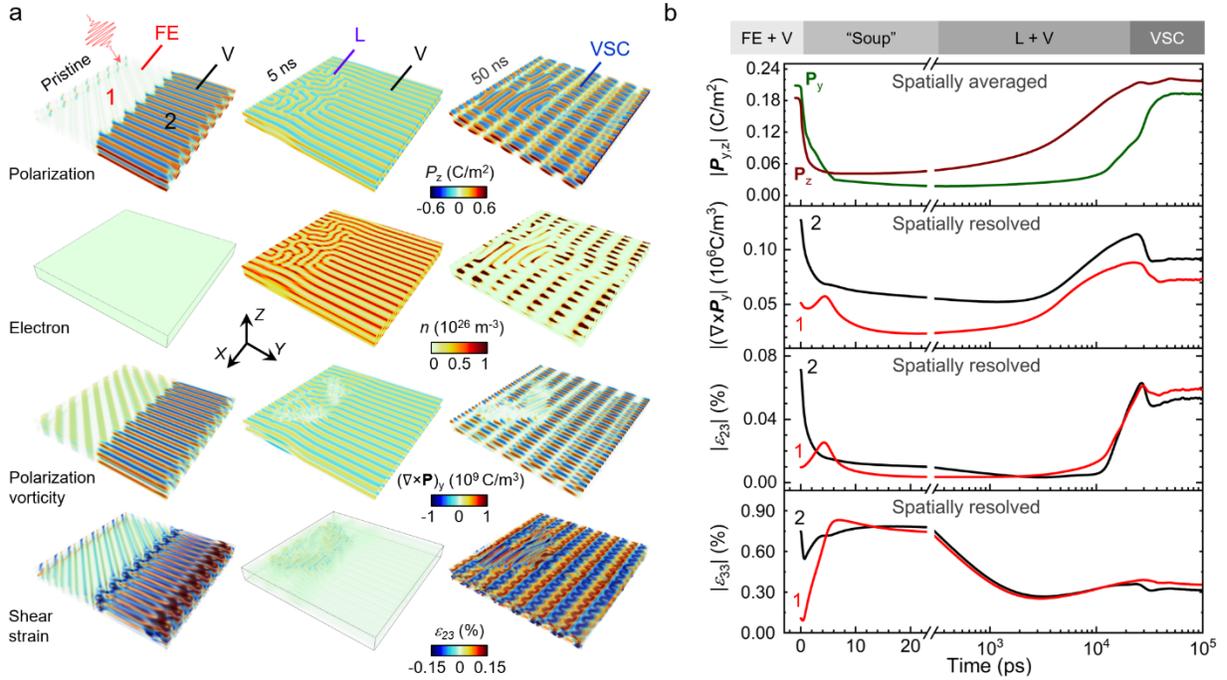

**Fig. 4 Simulated hierarchical order dynamics driving the vortex supercrystal formation after excitation with single optical pulses. a.** Spatiotemporal snapshots of polarization, electron concentration, polarization vorticity and shear strain evolving inside the simulated volume containing two PTO layers and two STO layers. Simulations demonstrate labyrinthine (L) fluctuations forming inside the original FE region and coexisting with V at 5 ns after single-shot optical excitation. At 50 ns, the VSC is formed and progresses towards a long-range order. **b.** The spatially averaged modulus of the polarization in the top panel shows a fast collapse of $P_z$ and $P_y$ components of the polarization on the ps timescale. The $P_z$ recovery starts at > 3 ns, while the $P_y$ recovery initiates at > 10 ns. The recovery of polarization vorticity (second panel) is captured separately in regions of the sample containing FE- (red curve, labeled 1) and V- (black curve, labeled 2) and phases. The polarization vorticity starts to grow in both regions at ~ 3 ns synchronous with the recovery of spatially averaged $P_z$. The shear strain, $\varepsilon_{23}$ (third panel) captures the interplay between these quantities and the polarization dynamics along the pathway to the VSC. In the FE region, a transient peak appears synchronously in polarization vorticity and $\varepsilon_{23}$ at ~ 5 ps, signaling the collapse of FE and formation of spatially incoherent reorientations of polarization from in-plane to out-of-plane. The shear strain, $\varepsilon_{23}$, starts to suddenly growth at > 10 ns, at the same time with the onset of $P_y$ recovery, also coinciding with VSC nucleation. In the fourth (bottom) panel, the normal strain, $\varepsilon_{33}$, is quantitatively consistent with the relative strain difference of V- and FE- phases before excitation, while this difference vanishes in ~ 10 ps. A non-monotonous evolution of $\varepsilon_{33}$ occurs between 3 ns and ~ 30 ns and can be correlated with the recovery of polarization vorticity and averaged $P_z$, revealing the normal strain signature of VSC formation.

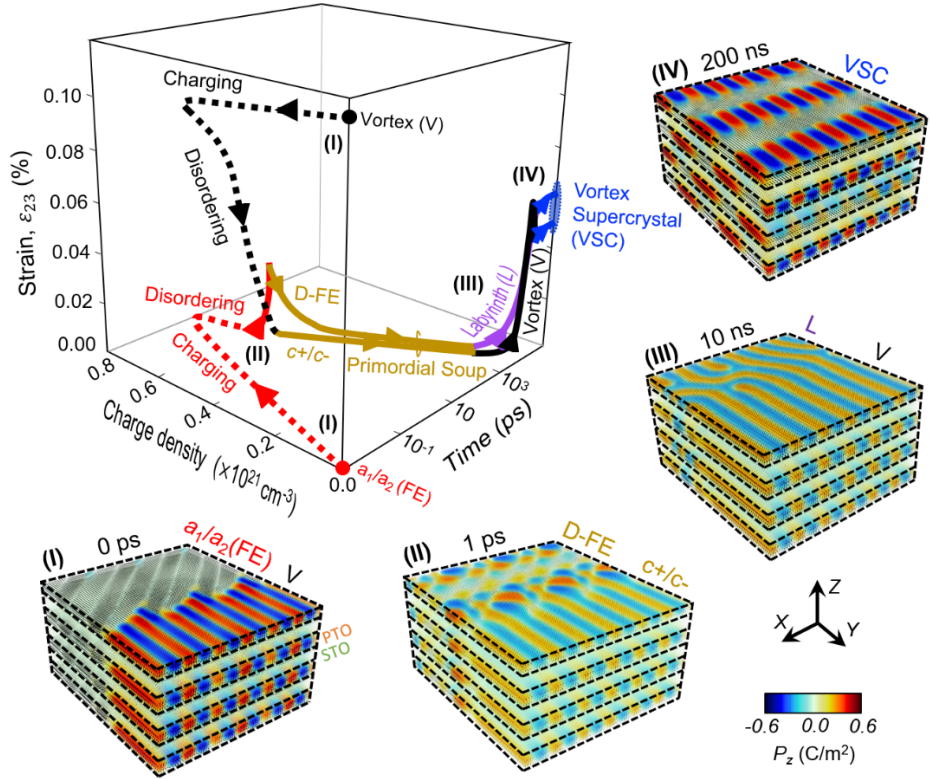

**Fig. 5 Three-dimensional overview of spatially resolved phase evolution from DPFM simulations.** A dynamical phase-field model (DPFM) considers the optical excitation through experimentally estimated photocarriers and thermal transients, using their relaxation rates to demonstrate the pathway of phase conversion, which is in favorable agreement with experiments. The three-dimensional plot captures the pathway of nonvolatile phase transition during the time evolution of shear strain, $\varepsilon_{23}$, and charge density dynamics in the spatial regions where the initial phases (V and FE) reside. At representative time delays, the simulated spatial distribution of polarization inside the superlattice overlays local polarization vectors with false color representing the $P_z$ component of polarization. The regime of sub-ps non-equilibrium disordering of V- and FE- phases observed in experiments is indicated by dotted lines. After subsequent thermalization of the system, on a few ps timescale, the V-phase converts to $c^+/c^-$, while the FE converts to a disordered ferroelectric phase (D-FE); their mixture is dubbed the "primordial soup". Starting within one nanosecond and until VSC nucleates at > 10 ns, labyrinthine fluctuations gradually replace D-FE, while the V phase reemerges from the "soup". At the longest time delays (> 20 ns), L and V merge into the VSC-phase, mediated by ferroelastic interactions at their phase boundaries exhibiting a sudden increase in $\varepsilon_{23}$ in both regions of the sample.

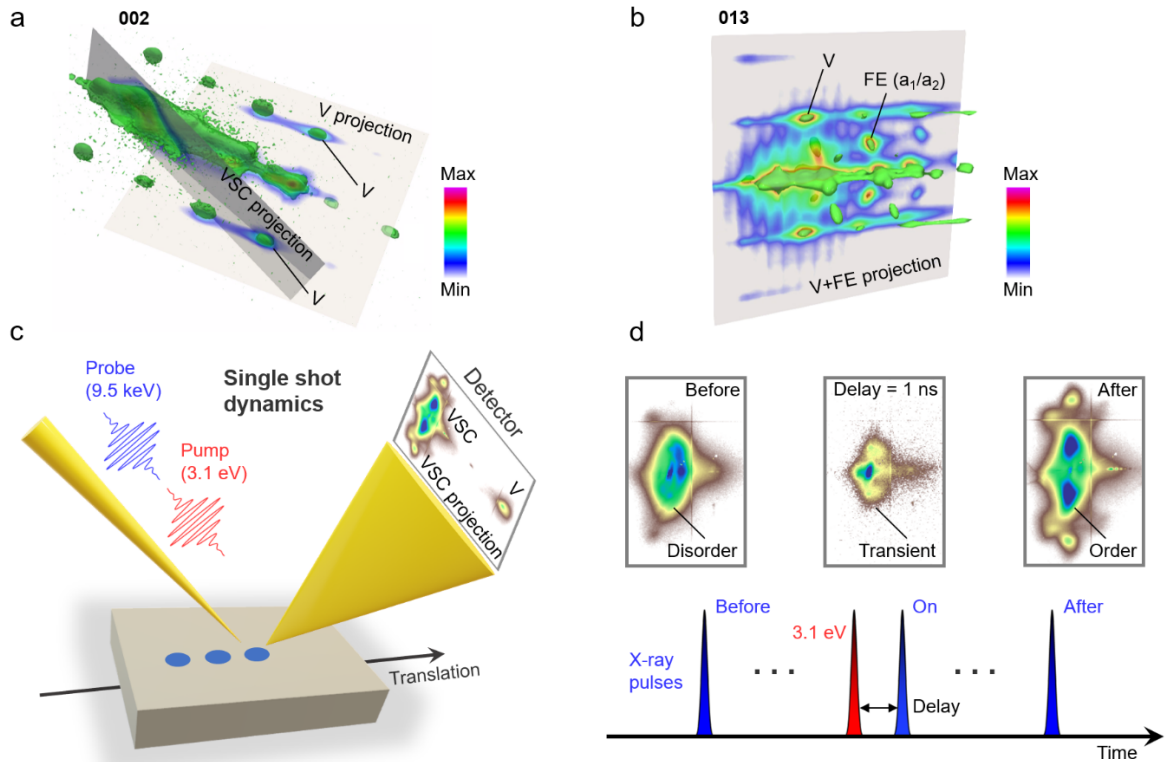

**Extended Data Fig. 1 Diffraction conditions and single-shot measurement protocol at X-ray free electron laser. a.** Selected detector planar projections relative to the diffraction pattern around the 002 Bragg peak measured at synchrotron in a $[(SrTiO_3)_{16}/(PbTiO_3)_{16}]_{7.5}$ superlattice sample. The VSC detector projection intersects the position of the emerging VSC satellite peaks and a V satellite peak, which is used in Fig. 1a for time-dependent measurements. The V detector projection intersects the V satellite peaks, which is used in Fig. 3a for time-dependent measurements. **b.** The detector planar projection relative to the diffraction pattern around the 013 Bragg peak measured at synchrotron in a $[(SrTiO_3)_{16}/(PbTiO_3)_{16}]_{7.5}$ superlattice sample. The V+FE detector projection intersects the position of emerging VSC satellite peaks and a V satellite peak, which is used in Fig. 2a for time-dependent measurements. **c.** Measurement geometry for $[(SrTiO_3)_{16}/(PbTiO_3)_{16}]_{7.5}$ superlattice samples. Each sample is translated to expose fresh regions, while a selected diffraction condition is captured by an area detector using the diffraction geometry shown in a. **d.** The single-shot time resolved measurement protocol involves recording diffraction patterns from 20-30 shots at 30 Hz rate "before" and "after" the laser pulse arrival. The single-shot transient diffraction pattern ("on" shot) is captured on sub-ps to µs timescales after the laser pulse at the diffraction condition shown in a. The image captured at 1 ns is compared to "before" and "after" images.

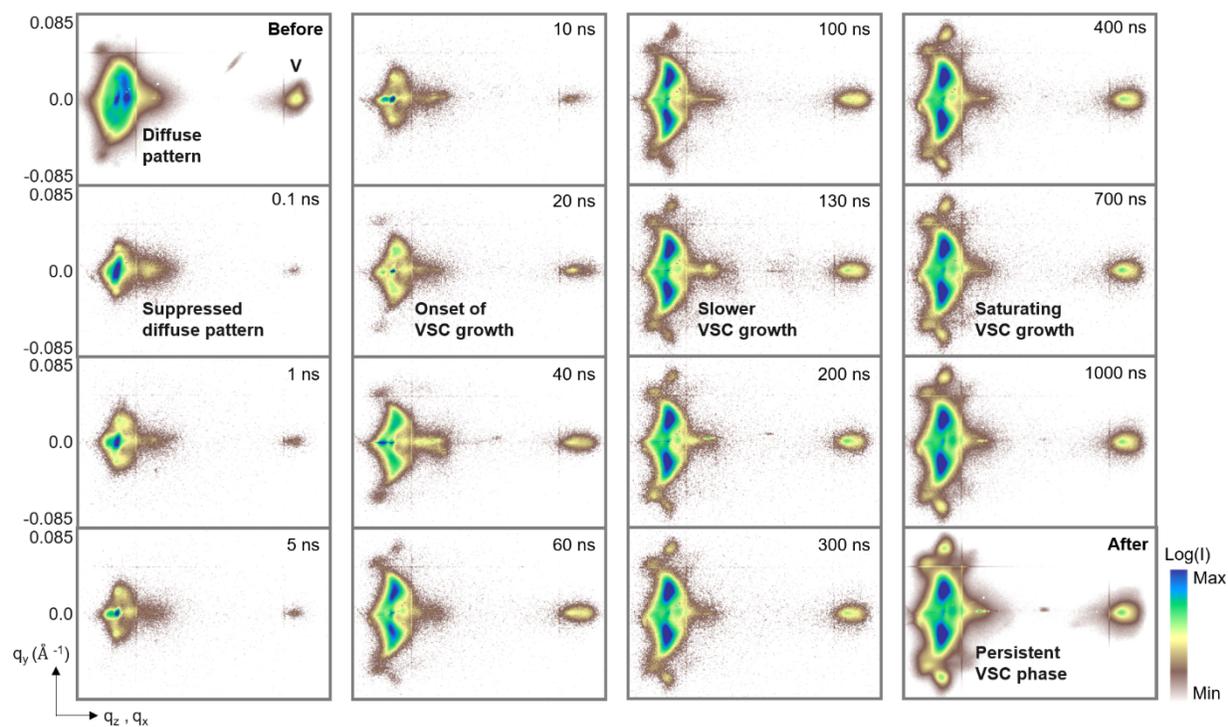

**Extended Data Fig. 2 Single-shot dynamics of vortex supercrystal formation measured near the 002 Bragg peak.** The detector planar surface, intersecting the V satellite peak (shown in Extended Data Fig. 1a) and VSC peaks (shown in the after images from Extended Data Fig. 1d), probes a detailed temporal evolution of the nonvolatile phase transformation in a $[(SrTiO_3)_{16}/(PbTiO_3)_{16}]_{7.5}$ superlattice sample. The diffraction geometry probes an extended time dependence of VSC growth, where selected time delays are marked in the inset for changes in the dynamical behavior. The logarithmic scale (false color) is kept the same in all individual detector images after photon count rate normalization to the X-ray beam intensity monitor. The vertical position in the images shows $q_y$ values relative to the image center, which intersects the 002 Bragg peak. The horizontal position in the images mixes the $q_x$ and $q_z$ components of the peaks. Different regimes captured during the dynamical evolution of phase transformations are marked.

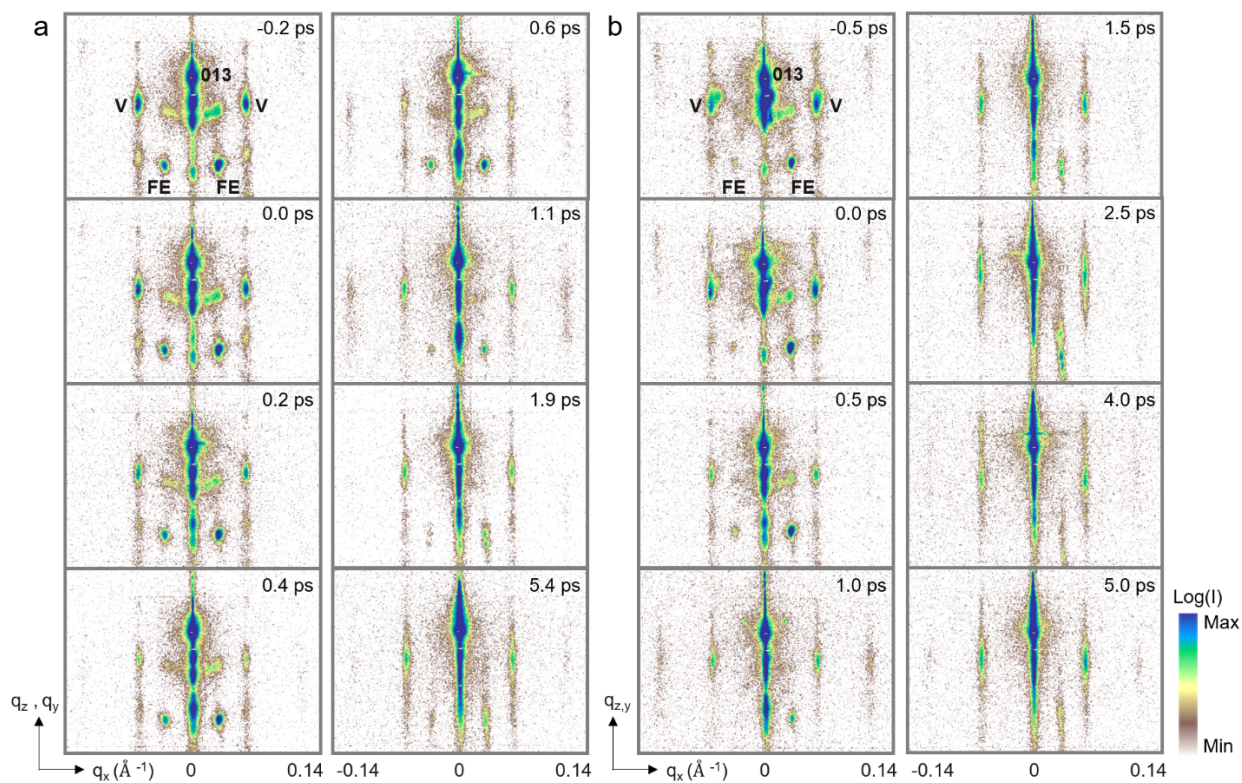

**Extended Data Fig. 3 Single-shot dynamics near the 013 Bragg peak.** Complementary to Fig. 2a, the dynamics at successive time delays are shown for two $[(SrTiO_3)_{16}/(PbTiO_3)_{16}]_{7.5}$ superlattice samples in **a.** and **b.**, respectively. The detector planar surface intersects V and FE satellite peaks allowing for direct comparison of their dynamics. The logarithmic false color scale is the same in all individual detector images after photon count rate normalization to the X-ray beam intensity monitor. The diffraction geometry is the same as in Fig. 2a and Extended Data Fig. 1b, monitoring the extended time dependence of V and FE satellite peaks. The horizontal position in the images is calibrated against the $q_x$ values relative to the crystal truncation rod, which intersects the 013 Bragg peak. The vertical position in the images mixes the $q_y$ and $q_z$ components of the peaks.

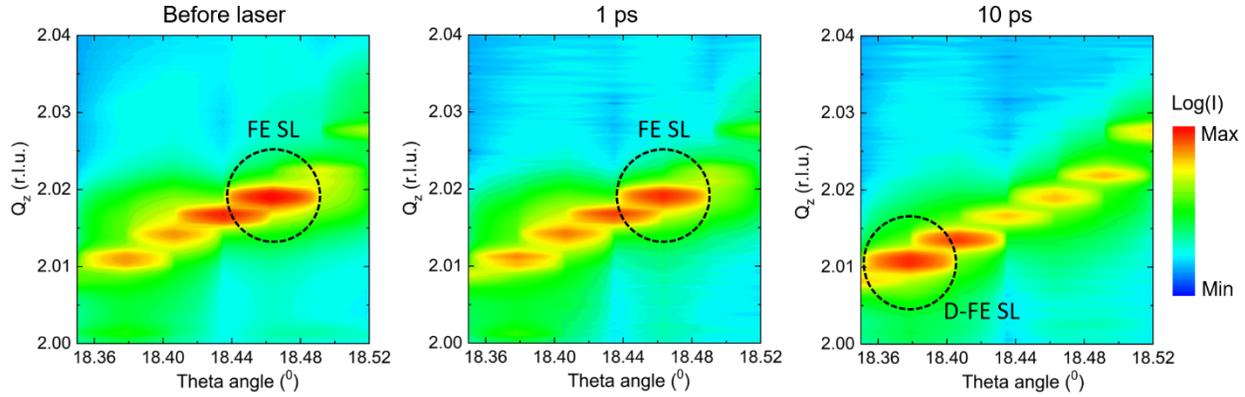

**Extended Data Fig. 4 Capturing the picosecond strain reconfiguration in the FE phase.**

False color maps of scattering X-ray intensities include line profiles along $q_z$ projection on the detector at a few representative angle of incidence values. The comparison of the map collected before laser excitation and the corresponding map at 1 ps shows that the FE-phase peak position does not change on this timescale. On the other hand, the map recorded at 10 ps shows that the FE peak moved to significantly smaller $q_z$, which is consistent with a tetragonality increase in this region of the sample. At 10 ps, the satellites of FE vanished (not shown), and the "primordial soup" emerged. The shifted superlattice peak shows that the polarization disorder persists throughout all layers of the superlattice.

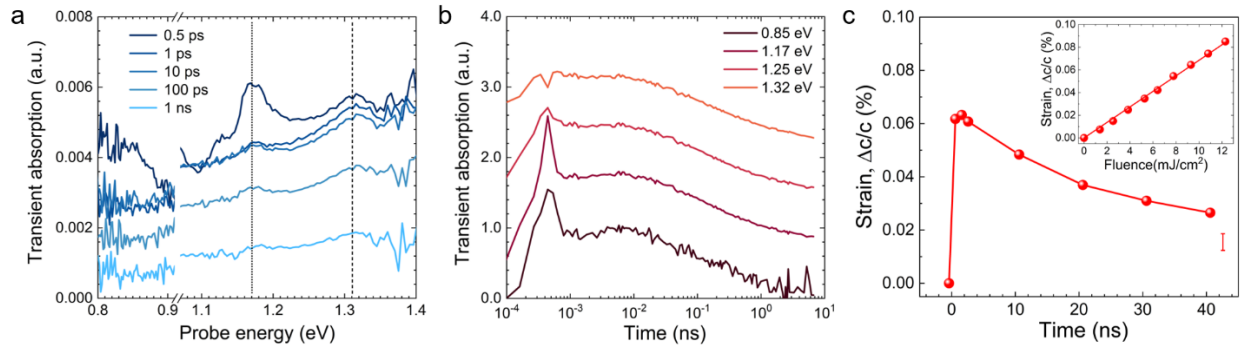

**Extended Data Fig. 5 Charge carrier and strain relaxation dynamics measured with stroboscopic pump-probe in a $[(SrTiO_3)_{16}/(PbTiO_3)_{16}]_{7.5}$ superlattice sample. a.** Transient absorption at a few representative time delays; two characteristic peaks (described in the text) are marked by the vertical lines. **b.** The temporal dynamics measured at the indicated energies for TA probe, which provide the charge carrier recombination rate used in DPFM simulations. **c.** The reversible temporal dynamics and fluence dependence measured at the synchrotron for the FE peak used to estimate the maximum temperature jump (~ 300 K) and thermal relaxation rate in DPFM simulations.

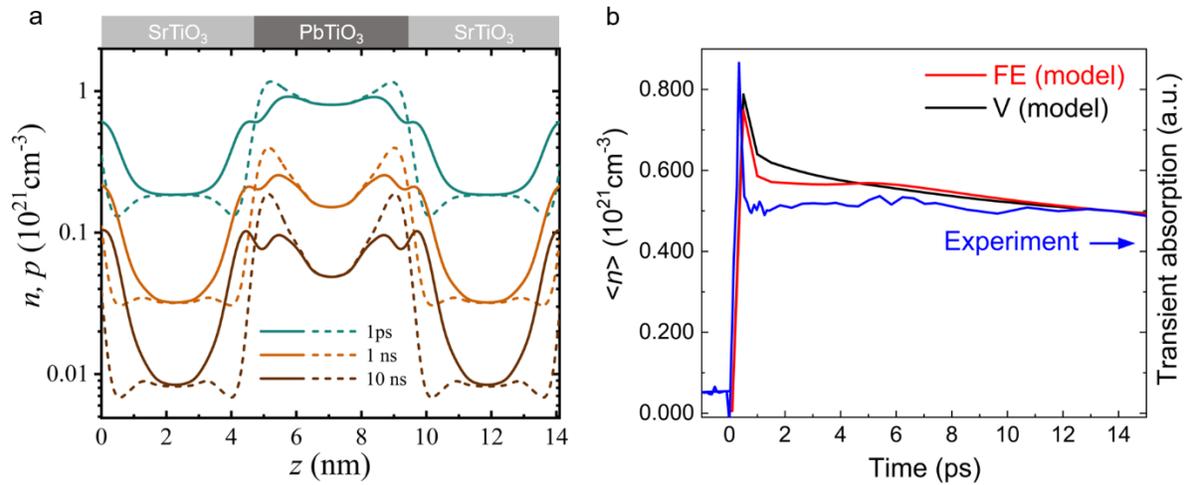

**Extended Data Fig. 6 Simulated charge-carrier dynamics showing the charge transfer at the superlattice interfaces. a.** In-plane spatially averaged electron ($n$, continuous lines) and hole ($p$, dotted lines) concentrations are plotted as function of vertical direction of the superlattice revealing a larger $n$ in SrTiO$_3$ layers near interfaces, while the larger $p$ is found in PbTiO$_3$ near the interfaces. **b.** The calculated dynamics of electrons in the two spatial regions of the sample are plotted against the experimental TA measurement.

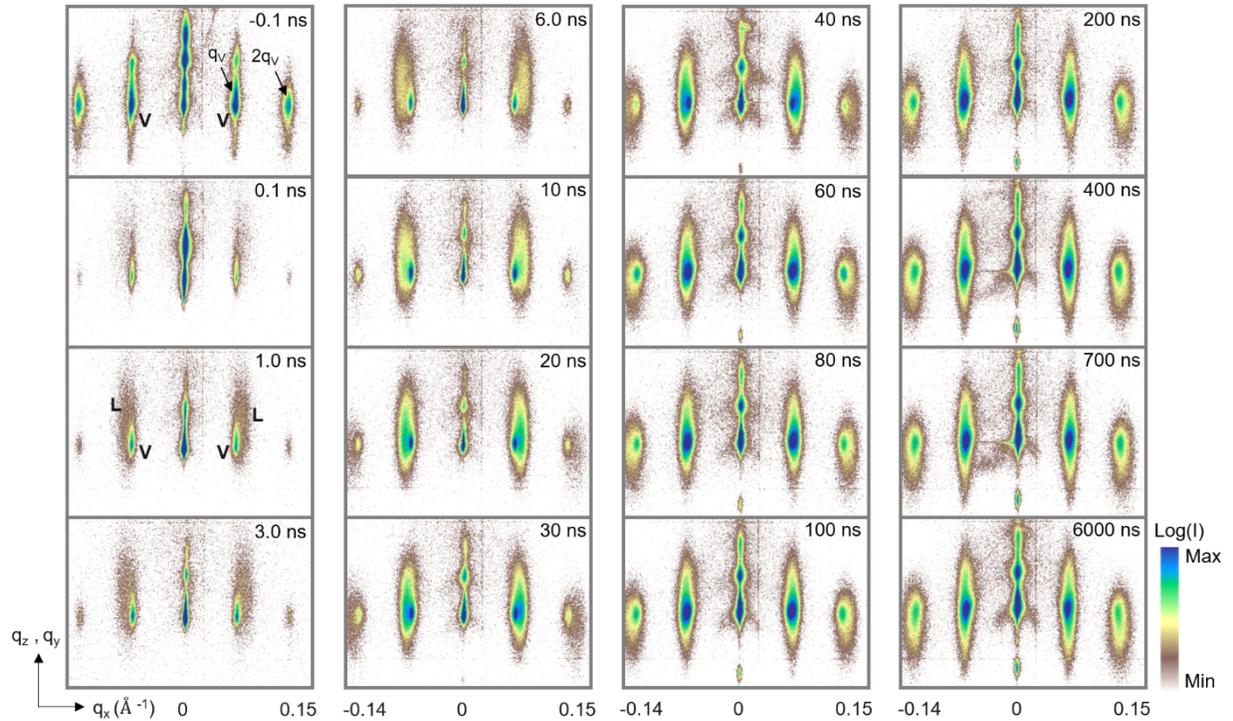

**Extended Data Fig. 7 Transient phases captured in single-shot dynamics near the 002 Bragg peak.** The detector planar surface probes the dynamics at successive time delays as indicated in a $[(SrTiO_3)_{16}/(PbTiO_3)_{16}]_{7.5}$ superlattice sample, intersecting the first and second order satellite peaks of the V-phase, which are indicated by arrows. The diffraction geometry is the same as in Figs. 3a and Extended Data 1a. A transient labyrinthine (L) phase is observed and marked at time delays of 0.1-10 ns. The appearance of second-order peaks at > 20 ns signals the simultaneous conversion of L- and V- phases to the VSC-phase. The vertical position in the images mixes the $q_y$ and $q_z$ components of the peaks.

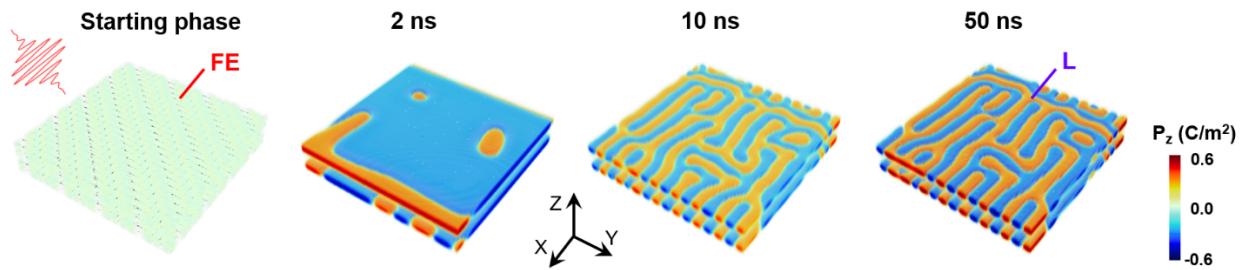

**Extended Data Fig. 8 Model for dynamics of FE transformation to L phase after excitation with single optical pulses.** Spatiotemporal snapshots of polarization contrast evolving inside the simulated volume of a pure FE starting phase with DPFM, which contains two PTO layers separated by an STO layer. Calculations demonstrate labyrinthine (L) fluctuations forming between 2 and 10 ns after single-shot optical excitation, while a transition to VSC is not observed.

31